%% file: conference_101719.tex
\documentclass[conference]{IEEEtran}
\IEEEoverridecommandlockouts
\usepackage{cite}
\usepackage{amsmath,amssymb,amsfonts}
\usepackage{algorithmic}
\usepackage{standalone}
\usepackage{myPlotStyle}
\usepackage{booktabs}
\usepackage{url}
\usepackage{graphicx}
\usepackage{textcomp}

\usepackage{xcolor}
\def\BibTeX{{\rm B\kern-.05em{\sc i\kern-.025em b}\kern-.08em
    T\kern-.1667em\lower.7ex\hbox{E}\kern-.125emX}}
\renewcommand{\baselinestretch}{0.94}
\begin{document}

\title{Toolbox for Developing Physics Informed Neural Networks for Power Systems Components \\ 
\thanks{This work was supported by the ERC Starting Grant VeriPhIED, funded by the European Research Council, Grant Agreement 949899.}}

\author{\IEEEauthorblockN{Ioannis Karampinis\IEEEauthorrefmark{1}, Petros Ellinas, Ignasi Ventura Nadal, Rahul Nellikkath, Spyros Chatzivasileiadis}
\IEEEauthorblockA{\textit{Department of Wind and Energy Systems
} \\
\textit{Technical University of Denmark (DTU)
}\\
Kgs. Lyngby, Denmark \\
{\IEEEauthorrefmark{1}iokar}@dtu.dk}
}
\maketitle
\begin{abstract}

This paper puts forward the vision of creating a library of neural-network-based models for power system simulations. Traditional numerical solvers struggle with the growing complexity of modern power systems, necessitating faster and more scalable alternatives. Physics-Informed Neural Networks (PINNs) offer promise to solve fast the ordinary differential equations (ODEs) governing power system dynamics. This is vital for the reliability, cost optimization, and real-time decision-making in the electricity grid. Despite their potential, standardized frameworks to train PINNs remain scarce. This poses a barrier for the broader adoption and reproducibility of PINNs; it also does not allow the streamlined creation of a PINN-based model library. This paper addresses these gaps. It introduces a Python-based toolbox for developing PINNs tailored to power system components, available on GitHub \url{https://github.com/radiakos/PowerPINN}. Using this framework, we capture the dynamic characteristics of a 9th-order system, which is probably the most complex power system component trained with a PINN to date, demonstrating the toolbox capabilities, limitations, and potential improvements. The toolbox is open and free to use by anyone interested in creating PINN-based models for power system components.

\end{abstract}



\section{Introduction}

The increasing integration of renewable energy sources (RES) into modern power systems has significantly altered system dynamics, reducing inertia and introducing non-linear behaviors due to inverter-based resources (IBR). These changes pose new challenges in ensuring system stability and reliability \cite{Erdiwansyah2021, Alimi2020}, demanding computational methods that are both accurate and scalable. Power system components exhibit dynamic behavior that can be mathematically represented using ordinary differential equations (ODEs), which describe the evolution of system states over time. Solving these equations efficiently in terms of accuracy and time is crucial for analyzing transient stability, frequency control, and fault responses.

Traditional simulation approaches primarily rely on numerical solvers such as Runge-Kutta methods to approximate the system dynamics governed by ODEs \cite{sauerandpi}. 
As power systems grow in complexity, conventional techniques struggle to meet the demand for fast and efficient simulations, particularly when dealing with high-order models or complex non-linear interactions.

On the one hand, Machine Learning (ML) techniques, particularly Neural Networks (NNs), have been explored to accelerate power system simulations by learning complex system behaviors. However, such
solutions are characterized by their high dependence on data to ensure generalizability and trustworthiness \cite{Hamilton2022}. 

A promising alternative is Physics-Informed Neural Networks (PINNs), which integrate physical laws into the learning process. By embedding ODEs directly into the network architecture, PINNs reduce data dependency while preserving physical accuracy 
\cite{lagaris, RAISSI2019686}. 
Recent studies have demonstrated the potential of PINNs in power system applications \cite{Huang2023}, including transient stability analysis \cite{stiasny2023transientstabilityanalysisphysicsinformed} and parameter \& state estimation  \cite{chertkov}. 
However, applying PINNs to an entire power system remains infeasible due to excessive training time and computational complexity. Instead, a more practical approach is to use PINNs for modeling individual power system components, which can then be integrated into conventional simulators \cite{integratingprevious} or incorporated into a dedicated PINN-based simulation engine, such as PINNSim \cite{pinnsim}. Our vision is for users to be able to easily create a library of PINN-based models, which they can then use with the simulator of their choice; this is similar to the model library of conventional simulation software.

Despite their potential, there is currently no standardized framework for defining, training, and evaluating PINNs for power system components. This lack of standardization hinders broader adoption and reproducibility. It also does not allow the streamlined creation of such a PINN-based model library. This paper addresses these gaps. We introduce a modular and automated framework that streamlines the PINN training process, enabling efficient and reproducible simulations of power system components. Our contributions can be summarized as follows:
\begin{enumerate}
    \item We introduce a modular, automated, and open-source framework to train PINNs for a wide range of power system components.
    \item We demonstrate the effectiveness of our approach by applying it to a 9th-order system consisted of a synchronous machine (SM) with an Automatic Voltage Regulator (AVR) and a Governor, which is probably the most complex power system component trained with a PINN to date. The code is open-source and available to anyone wishing to create a library of PINN models for their simulations.
    
\end{enumerate}

Looking ahead, we envision this framework as a foundation for a future library of PINN-based models, enabling seamless integration into power system simulations, similar to existing model libraries in traditional simulation tools.
\section{Physics-Informed Neural Networks}

Power system components exhibit dynamic behavior that can be mathematically represented using ODEs. These equations describe the evolution of state variables over time and are typically framed as initial value problems (IVPs). An IVP refers to finding a function $\mathbf{x}(t)$ solution to a differential equation subject to given initial conditions. The uniqueness of the solution is ensured by the Picard-Lindelöf theorem, assuming that the function $f$ governing the ODE is Lipschitz continuous. A system of ODEs can be expressed as:
\begin{equation}
\frac{d}{dt} \mathbf{x} = f (t, \mathbf{x}, \mathbf{u}, \mu), \quad \mathbf{x}(t_0) = \mathbf{x}_0.
\label{odes}
\end{equation}

where, the function 
$f$ governs the time evolution of the system state vector $\mathbf{x}(t;\mathbf{x}_0, \mathbf{u}, \mu)$, vector $\mathbf{u}$ the inputs of the system, vector $\mu$ consists of the system parameters to be determined and $ \mathbf{x}_0 $ denotes the initial conditions.

The goal is to find the function $\mathbf{x}(t;\mathbf{x}_0, \mathbf{u}, {\mu})$ that satisfies \eqref{odes}, providing the time-domain response of the system. This function will offer the unique solution along time for the given $\mathbf{x}_0, \mathbf{u}$ and $\mu$, which we refer to as a trajectory.

However, due to the nonlinear nature of the problem, it is impossible to derive the closed-form solution of $\mathbf{x}(t)$ and therefore we rely on approximations:
\begin{equation}
\hat{\mathbf{x}}(t; \mathbf{x}_0, \mathbf{u}, \mu) \approx \mathbf{x}(t; \mathbf{x}_0, \mathbf{u}, \mu) 
\label{eq:approximation}
\end{equation}

\subsection{Classical Ordinary differential equations solvers}
Classical numerical solvers, such as the Runge-Kutta methods, are widely used to solve ODEs. These methods approximate the solution trajectory over a specified time interval \( T \) by iteratively computing discrete updates based on derivative evaluations at small time steps. However, this process can be computationally expensive, particularly for complex systems with many state variables or stiff equations.

A promising alternative approach is the use of ML models, e.g. NNs, which can serve as function approximators.

\subsection{Neural Networks as function approximators}

The standard feed-forward Neural Networks are theoritically capable of appoximating any function \cite{hornik1991approximation}. NNs consist of multiple interconnected layers, each containing a number of nodes. These networks include a set of trainable parameters, namely the weights and biases of the connections between nodes, which are updated iteratively until they can approximate the target function with a given accuracy, with the help of an optimizer. Activation functions are applied at each node to introduce non-linearity, enabling the network to approximate complex, non-linear functions as in \eqref{eq:approximation}.

\subsection{Training Neural Networks with data}
NNs are typically trained using a dataset of input-output pairs (labeled data). The training process enables the network to generalize and predict outputs for unseen inputs, provided that the training data sufficiently represent the target domain. Through an iterative procedure, the trainable parameters of the NN are continuously updated to align their outputs with the desired ones. However, this procedure depends entirely on the training data's quality and availability. In our case, such data can be offered with the help of ODE solvers. By simulating a set of trajectories and structuring them into the required input-output format, we obtain the dataset necessary for training, validation, and testing of neural networks. Specifically, we construct a dataset of simulated trajectories within the input domain $\Omega_d$, consisting of a total of $N_d$ data points, represented as: $ \{ ({\mathbf{x}_{0}^{(i)}}, t^{(i)}), \mathbf{x}^{(i)} \}_{i=1}^{N_d} $. The Mean Squared Error (MSE) between the known state $\mathbf{x}^{(i)}$  and the NN output $ \mathbf{\hat{x}}^{(i)}$ for a given initial condition $(\mathbf{x}_{0},t)$, computed across
$N_d$ data points in the dataset, is given by the following loss term:
\begin{equation} \label{eq:data_loss}
\mathcal{L}_{\text{data}} = \frac{1}{N_d} \sum_{i=1}^{N_d} \left| \mathbf{\hat{x}}^{(i)} - \mathbf{x}^{(i)} \right|^2
\end{equation}

\subsubsection{Mathematical foundation of Physics Informed Machine Learning}
The fusion of well-established physical laws described by differential equations and the
capabilities of ML has offered a novel approach to modeling complex
physical systems. The proposed ML algorithms are characterized by their capability to combine data-driven and physics-driven solutions. The automatic differentiation(AD) function of deep learning models, such as NNs, provides the derivative of the predicted output with respect to a known input, e.g. time. In our case, this value represents the left-hand side of the ODE \eqref{odes}. For the right-hand side, the known output $\mathbf{x}$ from the aforementioned dataset is used as a reference value, e, while all other parameters remain known. As a result, we can introduce a physics-informed loss that enforces these two sides to be equal \eqref{eq:physics_data_loss}, ensuring that the trained ML model satisfies the underlying physics laws.
The physics-informed loss that is based on simulated data can be formed as follows:
\begin{equation} \label{eq:physics_data_loss}
\mathcal{L}_{\text{physics data}} = \frac{1}{N_d} \sum_{i=1}^{N_d} \left| \underbrace{\frac{d \mathbf{\mathbf{\hat{x}}}^{(i)}}{d t}}_{\text{ \tiny Derivative of NN output}} - \underbrace{f(t^{(i)}, \mathbf{x}^{(i)}, \mathbf{u}, \mu)}_\text{ \tiny Source term evaluation} \right|^2
\end{equation}

\textbf{Training Neural Networks without data}:
due to the nature of the ODEs, and the form of the above physics-informed loss, no labeled data are required for training. To enforce initial conditions, the neural network output must match the known values of the state variables at $t_0$. 
\begin{equation}
\label{eq:col_ic}
\mathcal{L}_\text{ic col} = \frac{1}{N_{ic}}\sum_{i=1}^{N_{ic}}\big|\mathbf{\hat{x}}^{(i)}(\mathbf{t}_0) - \mathbf{x}_0^{(i)}\big|^2
\end{equation}
where $N_{ic}$ is the number of distinct initial conditions considered
Additionally, the NN must satisfy the differential equation at arbitrary points within the input domain $\Omega$, leading to a physics-informed loss that does not require labeled data.
Conversely, using the NN output for the ODE's right-hand side creates a label-free physics loss. To compute this physics-based loss, we introduce collocation points, which are sampled points $\{(\mathbf{x_0}^{(i)}, t^{(i)}) \}$ within the input domain $\Omega_c$. These points allow us to evaluate the residuals of the ODE, ensuring that the NN adheres to the underlying physics. The physics loss is computed as:

\begin{equation} \label{eq:physics_col_loss}
 \mathcal{L}_{\text{physics col}} = \frac{1}{N_c} \sum_{i=1}^{N_c} \left| \underbrace{\frac{d \mathbf{\mathbf{\hat{x}}}^{(i)}}{d t}}_{\text{ \tiny Derivative of NN output}} - \underbrace{f(t^{(i)}, \mathbf{\hat{x}}^{(i)}, \mathbf{u}, \mu)}_\text{ \tiny NN approximation term evaluation} \right|^2
\end{equation}

where $N_{c}$ is the number of the collocation points considered.

\textbf{Hybrid training}

Training PINNs involves optimizing a loss function that combines both data-driven and physics-informed terms. These losses can originate from two sources: 
\begin{enumerate}
\item{Labeled Data Loss}, that ensures match between the NN output and simulated trajectories \eqref{eq:data_loss}, along with the corresponding physics-based penalty at these points \eqref{eq:physics_data_loss} 
\item{Collocation Loss}, which enforces the differential equation at thecollocation points where no labeled data are available \eqref{eq:physics_col_loss}, while also ensuring the NN satisfies the respective initial conditions \eqref{eq:col_ic}.
\end{enumerate}
The total loss function is then formulated as:
\begin{equation} \label{eq:total_loss}
\mathcal{L} = \lambda_d\mathcal{L}_{\text{data}}+\lambda_{dp}\mathcal{L}_{\text{data physics}}+\lambda_{cp}\mathcal{L}_{\text{col physics}} + \lambda_{ic}\mathcal{L}_{\text{ic col}}
\end{equation}
where $\lambda$ values balance contributions from data consistency and physics compliance. By minimizing 
$\mathcal{L}$, the network learns a solution that aligns with both empirical data and the governing equations, enhancing generalization in physics-informed learning.

\section{Training PINNs - Challenges}

The primary objective of this work is to develop an efficient and robust pipeline for training PINNs to simulate various dynamic components, accompanied by the corresponding Python code. The pipeline is designed to be fully parameterizable and easily customizable at every step. In the following sections, we outline the key stages of the methodology and provide a detailed justification for the approach taken in each stage.


\subsection{Setting up the ODEs}
The first step is to define the ODEs that govern the examined power system dynamic component without replacing any value. The inputs of the system $\mathbf{u}$ and the system parameters $\mu$ in \eqref{odes} are considered static and are stored separately. 
In that way, it is easy to test the power system components across a range of values for each parameter, e.g. different inertia for a synchronous machine.
These ODEs will be utilized by the ODE solver for dataset generation and for the loss function formulation during the NNs training. 

\subsection{Generating the Dataset}
The next step is to generate a dataset of simulated trajectories and one of collocation points, which will be used in training and testing phase of the PINN model. If real sensor data are available for the entire examined system, a portion of the simulated trajectories can be disregarded.

\textbf{Sampling the initial conditions}: 
We aim to approximate the behavior of a system within a specific input domain $\Omega_d$. We start by choosing the bounds of that domain, and also the number of different samples of each state. Different sampling methods are available, such as random, linear or \textbf{Latin Hypercube sampling} (LHS) \cite{mckay2000samplingcomparison}. LHS is preferred due to its ability to provide comprehensive coverage of the input space with fewer samples. This approach generates a diverse set of initial conditions for the system. Similarly, we will sample the initial conditions for the collocation points.

\subsection{Generating the trajectories}
As previously discussed, obtaining a closed-form solution is not feasible, necessitating the use of numerical methods to approximate the solution. Given the stiffness of the power system's equations, the higher-order \textbf{Runge-Kutta 4-5} method provides satisfactory accuracy with a computational cost that is negligible during the training phase. 
Subsequently, defining the \textbf{time horizon} and the \textbf{number of points} within the time interval enables the generation of the desired number of trajectories. 

\subsection{Preprocessing the Dataset}
For given initial conditions and the time $t$, NNs must approximate the state of the system at that time. As a result, it is essential to preprocess the generated trajectories to create this input-output format. These labeled data are used during training through the respective labeled losses \eqref{eq:data_loss},\eqref{eq:physics_data_loss} and for testing the NNs. For the collocation points, the desired input format is created by pairing each sampled initial condition with all the corresponding time points within the domain.

Additionally, it was observed that skipping certain points from the same trajectories 
significantly accelerate the training process while not compromising the model accuracy by using fewer points per trajectory. In other words, to maintain a high model accuracy it is important to have as many trajectories as possible (i.e. a broad range of initial conditions), but it is not necessary to have too many points along each trajectory.


\subsection{The architecture of Neural Network}

\begin{figure}[t]
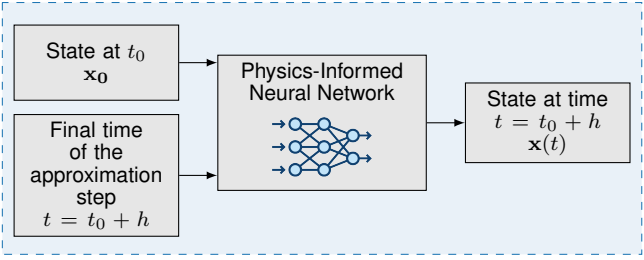

\centering
\includestandalone{Plots/NNs/nn_architecture}
\caption{Architecture of the proposed PINN that approximates the state \( \mathbf{x}(t) \) at time \( t = t_0 + h \) based on the initial state \( \mathbf{x}_0 \) and the time t.}
\label{fig:nn_architecture}
\end{figure}

The complexity of the examined power system component directly affects the architecture of the developed NN. The more complex this system is, the more neurons and/or layers must be included. As for the input layer, it must have K+1 inputs and K outputs as presented in Fig.~\ref{fig:nn_architecture}, where K is equal to the different dynamic states of the examined system. The extra input expresses the time for which the NN will offer the predicted state. Additionally, initialization weights strategies such as xavier can be included in order to promote stable NN training. Based on the accuracy and time inference needs, ML models can be implemented, such as Kolmogorov-Arnold
Networks (KANs) \cite{Ellinas2024}.

\subsection{Training challenges of the Neural Networks}

The training of the PINNs stands out as the most intricate process in this study, which
encompasses numerous components and a multitude of hyperparameters. Extensive experimentation was conducted to refine the training process and its hyperparameters. For training, the MSE loss function was selected due to its sensitivity to larger errors, which helps to effectively minimize significant deviations during the learning process. Regarding optimizers, various options were explored, with Adam and LBFG-S demonstrating superior performance, and a fixed
learning rate is preferred. In addition, early stopping can be integrated into the training pipeline to prevent overfitting. 

The choice of loss function weights $\lambda$ in \eqref{eq:total_loss} plays a crucial role in the efficient training of PINNs and can be either static or dynamic. To determine optimal initial weight magnitudes, we first perform a preliminary training phase using only $\mathcal{L}_{\text{data}}$. This step helps identify proper weight magnitudes, ensuring robust optimization and improved overall performance. In the actual training phase, a dynamic weighting strategy can be applied, where the weights of each loss component are adjusted based on either the training epoch or a strategy that considers the convergence behavior of individual loss terms. Inspired from \cite{heydari2019softadapt}, we explored a soft-adaptive weighting method, which falls under gradient-based approaches and dynamically adjusts loss weights based on their optimization dynamics. However, since this method resulted in comparable weight scaling and only slightly improved overall performance, we report results primarily using the static approach for clarity and conciseness.

\subsection{Performance assessment}
The developed PINN model will be evaluated in two distinct categories, accuracy and computation time:

\textbf{Accuracy}: A subset of the simulated data will be used as ground truth data. We calculate the MSE, the Mean Absolute Error (MAE), and the Maximum Absolute Error (Max AE) to evaluate the PINN's performance. MAE provides a linear measure of the average magnitude of errors. The Max AE measures the largest absolute error, drawing attention to the model's worst-case prediction performance. 

\textbf{Computation time}: In terms of time efficiency for approximating trajectories, it is crucial to highlight the reduced computation time required by PINNs compared to traditional numerical methods. As a benchmark, the time needed to obtain the solution for a single set and 100 sets of initial conditions is examined. This metric aims to demonstrate the efficiency of PINNs and their ability to parallelize computations effectively, making them well-suited for real-time applications and computationally demanding tasks.

Additionally, visualizations of the solutions and benchmarks can be highly valuable, providing insights into potential weaknesses of the proposed PINN model, particularly in approximating stiff trajectories.

\section{Experimental evaluation - Demonstration}

We apply the proposed methodology to a 9th- order system: a SM model equipped with an AVR and a Governor connected to an Infinite-Bus system. The following sections provide a detailed description of the system, including the parameters of the simulated dataset and the NN. Additionally, we outline the training setup along with its associated hyperparameters.

\subsection{CASE STUDY: The 4th-order Synchronous Machine with Automatic Voltage Regulator and Governor}

This section briefly presents the examined power system component. 
This study considers a 4th-order SM with an AVR and a governor connected to an infinite bus. The AVR and governor are represented by 3rd-order and 2nd-order models, respectively, thereby increasing the overall model complexity. 
As a result, the final system will be a 9th-order system, with the following representation found in \cite{sauerandpi}:

\begin{multline}
\hspace{-1.2em}
\scriptsize
\begin{bmatrix}
\scriptsize
1 \\
\frac{2H}{\Omega_{B}} \\
T'_{do} \\
T'_{qo} \\
T_E \\
T_F \\
T_A \\
T_{CH} \\
T_{SV}
\end{bmatrix}
\hspace{-0.2em}\frac{d}{dt} \hspace{-0.2em}
\begin{bmatrix}
\delta \\
\omega \\
E'_q \\
E'_d \\
E_{fd} \\
R_f \\
V_R \\
P_M \\
P_{SV}
\end{bmatrix}
\hspace{-0.2em}= \hspace{-0.2em}
\begin{bmatrix}
\scriptsize
\omega \\
P_m - E'_d I_d - E'_q I_q - (X'_q - X'_d)I_d I_q - D\omega \\
-E'_q - (X_d - X'_d)I_d + E_{fd} \\
-E'_d + (X_q - X'_q)I_q \\
-\left(K_E + S_E(E_{fd})\right) E_{fd} + V_R \\
-R_f + \frac{K_F}{T_F} E_{fd} \\
K_A R_f - \frac{K_A K_F}{T_F} E_{fd} + K_A \left(V_{\text{ref}} - V_t\right)  -V_R \\
-P_M + P_{SV} \\
-P_{SV} + P_C - \frac{1}{R_D} \left(\frac{\omega}{\Omega_B}\right)
\end{bmatrix}
\end{multline}
and 
\begin{align}
\begin{bmatrix}
(R_s + R_e) & -(X_q + X_e) \\
(X'_d + X_e) & (R_s + R_e)
\end{bmatrix}
\begin{bmatrix}
I_d \\ I_q
\end{bmatrix}
&=
\begin{bmatrix}
V_s \sin(\delta - \theta_{vs}) \\
V_s \cos(\delta - \theta_{vs}) \\
\end{bmatrix}
\end{align}
\begin{align}
\begin{bmatrix}
V_d \\ V_q \\ V_t \\ S_E(E_{fd})
\end{bmatrix}
&=
\begin{bmatrix}
R_e I_d - X_{ep} I_q + V_s \sin(\delta - \theta_{vs}) \\ 
R_e I_q - X_{ep} I_d + V_s \cos(\delta - \theta_{vs}) \\ 
\sqrt{V_d^2 + V_q^2} \\
0.098 * e^{(0.55*E_{fd})})
\end{bmatrix}
\end{align}
\begin{align}
\begin{bmatrix}
V_R^{\min} \\
0
\end{bmatrix}
\leq
\begin{bmatrix}
V_R \\
P_{SV}
\end{bmatrix}
\leq
\begin{bmatrix}
V_R^{\max} \\
P_{SV}^{\max}
\end{bmatrix}
\end{align}

The parameters used for the SM modeling are as follows: the damping factor is \( D = 2 \),  the inertia constant is \( H = 5.06 \) s and the stator resistance is \( R_s = 0 \) p.u. The direct-axis transient time constants are \( T_d' = 4.75 \) s, respectively, whereas the quadrature-axis transient time constant is \( T_q' = 1.6 \) s. 

The direct-axis synchronous, transient reactances are \( X_d = 1.25 \) p.u., \( X_d' = 0.232 \) p.u., respectively. Similarly, the quadrature-axis synchronous transient  reactances are \( X_q = 1.22 \) p.u., \( X_q' = 0.715 \) p.u.
The line and system parameters are as follows: the reactance of the line between the SM and the bus is \( X_{ep} = 0.1 \) p.u., while the resistance of the line is \( R_e = 0 \) p.u. The angular frequency of the system is \( \Omega_b = 314.159 \) rad/s, and the system input voltage magnitude is set to \( V_s = 1 \) p.u. The voltage phase angle is \( \theta_{vs} = 0 \) rad.

As for the AVR, the following parameters were used: the gain of the regulator is \( K_A = 20 \), and its time constant is \( T_A = 0.2 \, \text{s} \). The feedback gain is \( K_F = 0.063 \), with a feedback time constant of \( T_F = 0.35 \, \text{s} \). The excitation system gain is \( K_E = 1.0 \), and the excitation system time constant is \( T_E = 0.314 \, \text{s} \). The reference voltage is set to \( V_{\text{ref}} = 1.095 \). The output of the AVR is constrained within the limits \( V_R^{\min} = 0.8 \) and \( V_R^{\max} = 8 \).
The parameters used for the governor modeling are as follows: the steady-state power command is \( P_c = 0.7  \,\text{pu} \), and the regulation droop is \( R_d = 0.05 \, \text{pu}\). The governor time constant is \( T_{CH} = 0.4 \, \text{s} \), and the servo time constant is \( T_{SV} = 0.2 \, \text{s} \). The governor output is limited by a maximum servo power value of \( P_{SV}^{\max} = 1.0  \,\text{pu} \).
\subsubsection{Input domain}
The next step is to define the input domain $\Omega_d$ from which we will sample different initial conditions for our simulations. The input domain is specified as follows: $\theta \in [-2, 2]$ rad, $\omega \in [-1, 1]$ rad/s, $E_d'$ = 0 pu, $E_q' \in [0.9, 1.1]$ pu, $R_F $ = 1 pu, $V_r $ = 1.105 pu, $E_{fd}$ = 1.08 pu, $P_{sv}$ = 0.7048 pu, and $P_m$= 0.7048 pu. 500 different sets of initial conditions were sampled within this state space with LHS sampling. The initial conditions for the collocation points were also sampled using the same input domain and sampling method.
\subsubsection{Obtaining the final datasets}
Next, we want to approximate the behavior of the SM under transient conditions, and therefore we will consider that the
solver was operated with a 1 ms timestep and a simulation period of 1 s. The classical RK45-solver as implemented in \texttt{scipy.integrate} will be deployed to generate the simulated trajectories. The subsequent processing of these outputs results in the set $\mathcal{X}_d$ represented in the input-output form $(\mathbf{x}_0,t),\mathbf{x} $, which contains $N_d$ = 500'000 data points. 

The initial conditions for collocation points are sampled with LHS independently from those used in the labeled data. However, both datasets share the same time instances as in $\mathcal{X}_d$ to maintain consistency in the temporal domain. 
The resulting second dataset $\mathcal{X}_c$ is represented in the input form $(\mathbf{x}_0,t)$, which contains $N_c$ = 500'000 data points. 
\begin{subequations}\label{eq:T1T2}
\begin{align}
\mathcal{X}_d & = \{(\mathbf{x}_0^{(i)}, t^{(i)}), \mathbf{x}^{(i)} \}_{i=1}^{N_{d}=500'000} \\
\mathcal{X}_c & =  \{ ({\mathbf{x}_0^{(i)}}, t^{(i)}) \}_{i=1}^{N_{c}=500'000}  
\end{align}
\end{subequations}

The dataset $\mathcal{X}_d$ is used for training, validation, and testing of the NNs through the respective data losses, with a trajectory distribution of 80\%, 10\%, and 10\%, respectively. In contrast, $\mathcal{X}_c$ is used exclusively for training purposes. If needed, training can be performed using only one of these datasets, depending on the specific requirements of the task. As mentioned earlier, for efficient training, we sample at regular intervals across the same trajectory. The sampling interval is 23 steps for data points and 19 steps for collocation points.

\subsection{Training the PINN}

A NN with four hidden layers and 64 nodes with the hyperbolic tangent (tanh) activation function in each layer is used to approximate the state of the above system. ML algorithms were implemented using PyTorch \cite{pytorch}. The LBFG-S optimizer \cite{LBFGS}  with a learning rate of 0.001 was used to update the trainable parameters within 750 epochs, using the MSE loss criterion. As for the weights of the losses, we used the following static weights: $\lambda_d = 1$, $\lambda_{dp} = 0.01$, $\lambda_{cp}= 0.001$ and $\lambda_{ic} = 0.01$. WandB \cite{wandb} was used to monitor and tune the hyperparameters, providing insightful information regarding the training and testing performance. All experiments were conducted on the High-Performance Computing (HPC) cluster at the Technical University of Denmark, utilizing a 16-core Intel Xeon 6226R processor with 256 GB of RAM and an NVIDIA V100 GPU with 16 GB of exclusive GPU memory. 


\subsection{Results}


In this section, we evaluate the performance of the implemented PINN models in approximating the behavior of a 9th-order system. The objective is to assess the generalizability of the proposed framework when applied to a higher-order system. To this end, we compare the performance of the PINN models against the ODE solver on a test set comprising 50 unique trajectories.

\begin{figure}[t]
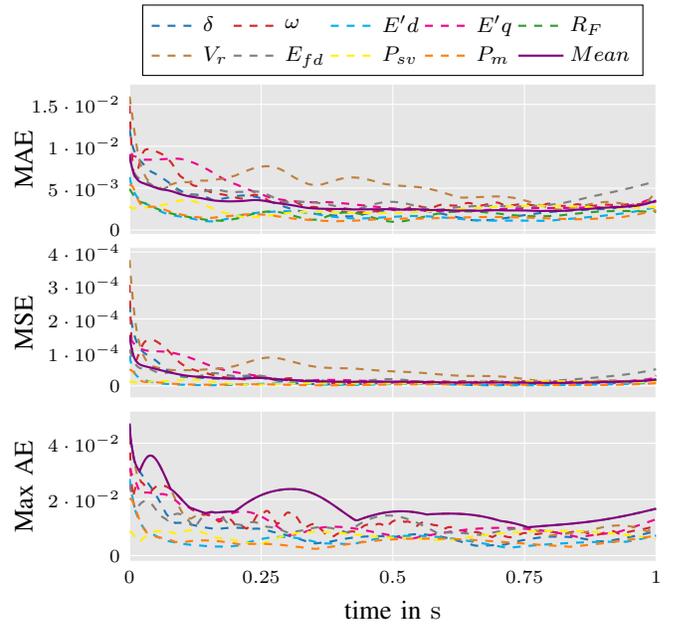

\centering
\includestandalone{Plots/accuracy_all}
\caption{Metric scores of a PINN approximating the states of a 9th-order SM. The error metrics are benchmarked against the ODE solutions.}
\label{fig:accuracy_results}
\end{figure}

In terms of accuracy, the results highlight the capability of PINNs to simulate the
dynamic behavior of higher-order system with 9 state variables. PINN approximated the states of the system with a MAE score below $2.26 \times 10^{-3}$ and Max AE $44.85 \times 10^{-3}$, which was noted in the very first moments according to Fig.~\ref{fig:accuracy_results}. 
Although this level of accuracy may be sufficient for certain applications, the current project has highlighted areas where further refinement is possible.

\begin{table}[t]
\centering
\caption{Accuracy of PINN}
\label{time_table}
\scriptsize
\centering
\setlength{\tabcolsep}{3pt} 
\renewcommand{\arraystretch}{0.9} 
\begin{tabular}{p{1.5cm}cc}
\toprule
 MAE & MSE & Max AE  \\
\midrule
 $2.26 \times 10^{-3}$ & $10.04 \times 10^{-6}$ & $44.85 \times 10^{-3}$ \\

\bottomrule
\end{tabular}
\end{table}

In terms of time efficiency, it is crucial to highlight the reduced computation time
required by PINNs compared to traditional numerical methods. 

Table~\ref{time_table} shows that as the inference state space grows, the advantages of PINNs become even more pronounced. Such approaches excel in their ability to parallelize outputs, allowing for efficient assessment of a broad range of initial parameters.

\begin{table}[t]
\centering
\caption{Inference time (in ms).}
\label{time_table}
\scriptsize
\centering
\setlength{\tabcolsep}{3pt} 
\renewcommand{\arraystretch}{0.9} 
\begin{tabular}{p{1.5cm}ccc}
\toprule
 Used Method & Single Trajectory & 50 Trajectories & 500 Trajectories  \\
\midrule
 ODE solver & 10.81  & 54.06  & 5406.13 \\
 PINN & 1.952 &  3.82 & 8.59 \\
\bottomrule
\end{tabular}
\end{table}

\begin{figure}[t]
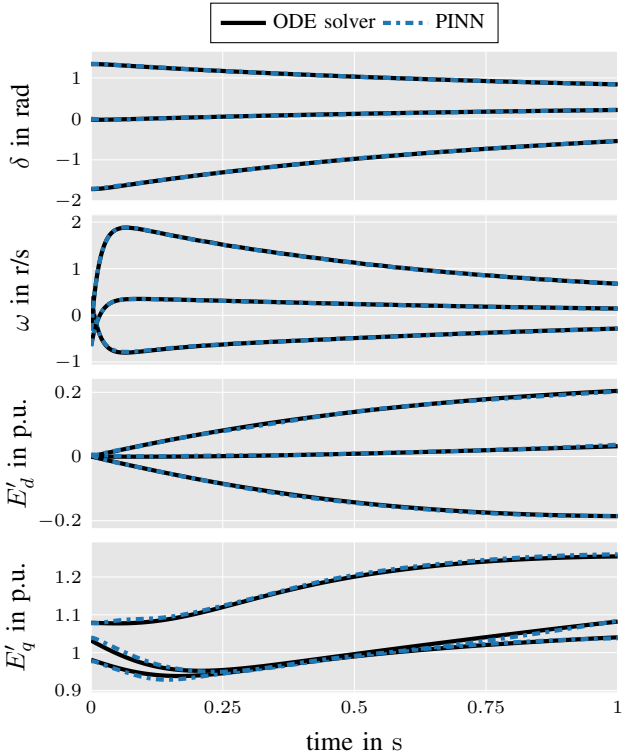


\includestandalone{Plots/trajectories}
\caption{Results for SM variables from the ODE solver and the trained PINN approximation for 3 different sets of initial conditions}
\label{fig:example_results}
\end{figure}

Finally, Fig.~\ref{fig:example_results} illustrates the trajectories of the state variables for the SM, generated using three randomly selected sets of test initial conditions. The results highlight that PINNs successfully capture the flow of the trajectories under varying initial conditions, adhering closely to the governing physical laws.




\section{Conclusion and Future Work}
This work developed an efficient, unified pipeline tailored for training PINN models on power system components and introduced a comprehensive toolbox featuring synchronous machines, designed to seamlessly integrate as plug-and-play modules.  The proposed approach demonstrated that PINNs can effectively approximate high-order system dynamics while significantly reducing computational costs compared to traditional solvers. Our test case showcased the ability of PINNs to model a 9th-order SM with AVR and Governor controllers, highlighting their potential for higher-order power system components.

Despite these advantages, several challenges remain. PINNs' accuracy depends on training data quality, loss function weighting, and hyperparameter selection. In addition, future work will explore adaptive sampling strategies and improve the selection of collocation points \cite{lau2024pinnacle} to enhance model accuracy and training efficiency. Expanding the component library to include emerging technologies such as inverter-based resources (IBRs) will further broaden their applicability. Additionally, integrating real-world measurement data with simulated datasets can improve robustness and generalization. Finally, incorporating PINNs into commercial simulation platforms, such as PowerFactory or PSCAD, will facilitate industrial adoption and practical deployment.

Overall, this study underscores the potential of PINNs in power system dynamic modeling and transient analysis while highlighting areas for further research to enhance their practical applicability.

\bibliographystyle{IEEEtran}
\bibliography{bibliography}

\end{document}

%% file: Plots/NNs/nn_architecture.tex
\begin{tikzpicture}[node distance = 3cm]
\sffamily
\footnotesize

\node[block](NN) {\begin{tabular}{c} Physics-Informed \\ Neural Network \\ [0.5em]
\scalebox{0.1}{\includegraphics[]{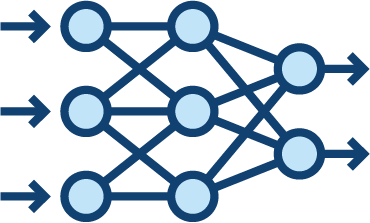}}\end{tabular}};

\node[block, left of = NN, text width = 2cm, align = center, yshift = 0.8cm, xshift = -1cm](x_in){State at $t_0$ \\ $\mathbf{x_0}$};
\node[block, below of = x_in, text width = 2cm, minimum width = 0.5cm, yshift = 0.5cm, align = center](t_in){Final time of the approximation step \\ $t=t_0+h$};

\node[block, right of = NN, text width = 2cm, align = center, xshift = 1cm](x_out){State at time $t = t_0+h$ \\
$\mathbf{x}(t)$};

\draw[-latex] (x_in.east) --(x_in.east-|NN.west);
\draw[-latex] (t_in.east) --(t_in.east-|NN.west);
\draw[-latex] (NN.east) --(x_out);

\begin{pgfonlayer}{bg}

    \path (x_in) -- +(-1.25,0.8) coordinate (r1);
    \path (x_out) -- +(1.25,-1.75) coordinate (r2);
    \draw[sBlue, dashed,fill = sBlue!10] (r1) rectangle(r2);
    \node [left of = r2, anchor = south east, node distance = 0.1, sBlue, yshift = 0cm] {
    };

\end{pgfonlayer}

\end{tikzpicture}

%% file: Plots/accuracy_all.tex
\begin{tikzpicture}
\begin{groupplot}[
    group style={
    group name = mygroup, 
    group size=1 by 3,          
    xticklabels at = edge bottom,
    vertical sep = 5pt
    },
    ylabsh=-4em,                   
    xlabel={},    
    xmin = 0, xmax = 1,    
    xtick distance = 0.25,         
    legend columns = 5,
    height=2cm,
    ylabel style={text width = 1.5cm, align = center}, 
    legend style={at={(0.5,1.05)}, anchor = south},
]
\nextgroupplot[
    ylabel={$\text{MAE}$},
    ]
    \plot[sBlue, dashed,  thick, ] table[x= t, y= theta]{Plots/SM_AVR_GOV_DynamicNN_mae.dat}; 
    \addlegendentry{$\delta$}  
    \plot[sRed, dashed, thick] table[x= t, y=omega]{Plots/SM_AVR_GOV_DynamicNN_mae.dat}; 
    \addlegendentry{$\omega$}  
    \plot[cyan, dashed, thick] table[x= t, y=E_d_dash]{Plots/SM_AVR_GOV_DynamicNN_mae.dat}; 
    \addlegendentry{$E'd$}  
    \plot[magenta, dashed, thick] table[x= t, y=E_q_dash]{Plots/SM_AVR_GOV_DynamicNN_mae.dat}; 
    \addlegendentry{$E'q$}  
    \plot[sGreen, dashed, thick] table[x= t, y=R_F]{Plots/SM_AVR_GOV_DynamicNN_mae.dat}; 
    \addlegendentry{$R_F$}  
    \plot[brown, dashed, thick] table[x= t, y=V_r]{Plots/SM_AVR_GOV_DynamicNN_mae.dat}; 
    \addlegendentry{$V_r$}  
    \plot[gray, dashed, thick] table[x= t, y=E_fd]{Plots/SM_AVR_GOV_DynamicNN_mae.dat}; 
    \addlegendentry{$E_{fd}$}  
    \plot[yellow, dashed, thick] table[x= t, y=P_sv]{Plots/SM_AVR_GOV_DynamicNN_mae.dat}; 
    \addlegendentry{$P_{sv}$}  
    \plot[sOrange , dashed, thick] table[x= t, y=P_m]{Plots/SM_AVR_GOV_DynamicNN_mae.dat}; 
    \addlegendentry{$P_m$}

    \plot[violet, thick] table[x= t, y=total]{Plots/SM_AVR_GOV_DynamicNN_mae.dat};   
    \addlegendentry{$Mean$}  
\nextgroupplot[
    ylabel={$\text{MSE}$ },
    ]
    \plot[sBlue, dashed,  thick, ] table[x= t, y= theta]{Plots/SM_AVR_GOV_DynamicNN_mse.dat}; 
    \plot[sRed, dashed, thick] table[x= t, y=omega]{Plots/SM_AVR_GOV_DynamicNN_mse.dat}; 
    \plot[cyan, dashed, thick] table[x= t, y=E_d_dash]{Plots/SM_AVR_GOV_DynamicNN_mse.dat}; 
    \plot[magenta, dashed, thick] table[x= t, y=E_q_dash]{Plots/SM_AVR_GOV_DynamicNN_mse.dat}; 
    \plot[brown, dashed, thick] table[x= t, y=V_r]{Plots/SM_AVR_GOV_DynamicNN_mse.dat}; 
    \plot[gray, dashed, thick] table[x= t, y=E_fd]{Plots/SM_AVR_GOV_DynamicNN_mse.dat}; 
    \plot[yellow, dashed, thick] table[x= t, y=P_sv]{Plots/SM_AVR_GOV_DynamicNN_mse.dat}; 
    \plot[sOrange , dashed, thick] table[x= t, y=P_m]{Plots/SM_AVR_GOV_DynamicNN_mse.dat}; 
    \plot[violet, thick] table[x= t, y=total]{Plots/SM_AVR_GOV_DynamicNN_mse.dat};   
   
\nextgroupplot[                     
    ylabel={$\text{Max AE}$ },         
    xlabel = time in \si{\second},
    ]
    
    \plot[sBlue, dashed,  thick, ] table[x= t, y= theta]{Plots/SM_AVR_GOV_DynamicNN_max_ae.dat}; 
    \plot[sRed, dashed, thick] table[x= t, y=omega]{Plots/SM_AVR_GOV_DynamicNN_max_ae.dat}; 
    \plot[cyan, dashed, thick] table[x= t, y=E_d_dash]{Plots/SM_AVR_GOV_DynamicNN_max_ae.dat}; 
    \plot[magenta, dashed, thick] table[x= t, y=E_q_dash]{Plots/SM_AVR_GOV_DynamicNN_max_ae.dat}; 
    \plot[brown, dashed, thick] table[x= t, y=V_r]{Plots/SM_AVR_GOV_DynamicNN_max_ae.dat}; 
    \plot[gray, dashed, thick] table[x= t, y=E_fd]{Plots/SM_AVR_GOV_DynamicNN_max_ae.dat}; 
    \plot[yellow, dashed, thick] table[x= t, y=P_sv]{Plots/SM_AVR_GOV_DynamicNN_max_ae.dat}; 
    \plot[sOrange , dashed, thick] table[x= t, y=P_m]{Plots/SM_AVR_GOV_DynamicNN_max_ae.dat}; 
    \plot[violet, thick] table[x= t, y=total]{Plots/SM_AVR_GOV_DynamicNN_max_ae.dat};   

\end{groupplot}
\end{tikzpicture}

%% file: Plots/trajectories.tex
\begin{tikzpicture}
\begin{groupplot}[
    group style={
    group name = mygroup, 
    group size=1 by 4,          
    xticklabels at = edge bottom,
    vertical sep = 5pt
    },
    ylabsh=-2.7em,                   
    xlabel={},    
    xmin = 0, xmax = 1,    
    xtick distance = 0.25,         
    legend columns = 4,
    height=2cm,
    legend style={at={(0.5,1.05)}, anchor = south}
]

\nextgroupplot[                     
    ylabel={$\delta$ in rad},         
    ]
    \plot[black, ultra thick] table[x= t, y=0]{Plots/samples/SM_AVR_GOV_y_true_samples_theta.dat};     
    \addlegendentry{ODE solver}    
     
    \plot[sBlue, dashdotted, ultra thick] table[x= t, y=0]{Plots/samples/SM_AVR_GOV_DynamicNN_pred_samples_theta.dat}; 
    \addlegendentry{PINN}
    

     
    
    \plot[black, ultra thick] table[x= t, y=2]{Plots/samples/SM_AVR_GOV_y_true_samples_theta.dat};    
     
    \plot[sBlue, dashdotted, ultra thick] table[x= t, y=2]{Plots/samples/SM_AVR_GOV_DynamicNN_pred_samples_theta.dat};     
    \plot[black, ultra thick] table[x= t, y=3]{Plots/samples/SM_AVR_GOV_y_true_samples_theta.dat};    
     
    \plot[sBlue, dashdotted, ultra thick] table[x= t, y=3]{Plots/samples/SM_AVR_GOV_DynamicNN_pred_samples_theta.dat};

     

\nextgroupplot[
    ylabel={$\omega$ in r/s},
    xlabel = {time in \si{\second}},
    ]
    \plot[black, ultra thick] table[x= t, y=0]{Plots/samples/SM_AVR_GOV_y_true_samples_omega.dat};     
     
    \plot[sBlue, dashdotted, ultra thick] table[x= t, y=0]{Plots/samples/SM_AVR_GOV_DynamicNN_pred_samples_omega.dat};     

     
    
    \plot[black, ultra thick] table[x= t, y=2]{Plots/samples/SM_AVR_GOV_y_true_samples_omega.dat};     
     
    \plot[sBlue, dashdotted, ultra thick] table[x= t, y=2]{Plots/samples/SM_AVR_GOV_DynamicNN_pred_samples_omega.dat};     
    \plot[black, ultra thick] table[x= t, y=3]{Plots/samples/SM_AVR_GOV_y_true_samples_omega.dat};     
     
    \plot[sBlue, dashdotted, ultra thick] table[x= t, y=3]{Plots/samples/SM_AVR_GOV_DynamicNN_pred_samples_omega.dat};

     
    
\nextgroupplot[
    ylabel={$E'_d$ in p.u.},
    ]
    \plot[black, ultra thick] table[x= t, y=0]{Plots/samples/SM_AVR_GOV_y_true_samples_E_d_dash.dat};     
     
    \plot[sBlue, dashdotted, ultra thick] table[x= t, y=0]{Plots/samples/SM_AVR_GOV_DynamicNN_pred_samples_E_d_dash.dat};   


    \plot[black, ultra thick] table[x= t, y=2]{Plots/samples/SM_AVR_GOV_y_true_samples_E_d_dash.dat};     
     
    \plot[sBlue, dashdotted, ultra thick] table[x= t, y=2]{Plots/samples/SM_AVR_GOV_DynamicNN_pred_samples_E_d_dash.dat};  
    
    \plot[black, ultra thick] table[x= t, y=3]{Plots/samples/SM_AVR_GOV_y_true_samples_E_d_dash.dat};     
     
    \plot[sBlue, dashdotted, ultra thick] table[x= t, y=3]{Plots/samples/SM_AVR_GOV_DynamicNN_pred_samples_E_d_dash.dat}; 

\nextgroupplot[
    ylabel={$E'_q$ in p.u.},
    xlabel = {time in \si{\second}},
    ]
    \plot[black, ultra thick] table[x= t, y=0]{Plots/samples/SM_AVR_GOV_y_true_samples_E_q_dash.dat};     
     
    \plot[sBlue, dashdotted, ultra thick] table[x= t, y=0]{Plots/samples/SM_AVR_GOV_DynamicNN_pred_samples_E_q_dash.dat};   
    \plot[black, ultra thick] table[x= t, y=2]{Plots/samples/SM_AVR_GOV_y_true_samples_E_q_dash.dat};     

    \plot[sBlue, dashdotted, ultra thick] table[x= t, y=2]{Plots/samples/SM_AVR_GOV_DynamicNN_pred_samples_E_q_dash.dat};  
    \plot[black, ultra thick] table[x= t, y=3]{Plots/samples/SM_AVR_GOV_y_true_samples_E_q_dash.dat};     
     
    \plot[sBlue, dashdotted, ultra thick] table[x= t, y=3]{Plots/samples/SM_AVR_GOV_DynamicNN_pred_samples_E_q_dash.dat};


\end{groupplot}
\end{tikzpicture}